# A CONCISE RESOLUTION TO THE TWO ENVELOPE PARADOX

ERIC BLISS

**Abstract**

In this paper I will demonstrate a new perspective on the Two Envelope Problem. I hope to clearly show how the paradox results from a problem pertaining to the interpretation of Bayesian probability when there is a subjective probability that is inconsistent with reality. Specifically, the results are paradoxical because the player on average is off by a larger amount when they expect a larger payoff that is not an actual possibility, than when he expects a smaller payoff that is not an actual possibility.

**Keywords**

Two Envelope Paradox, Two Envelope Problem, Bayesian, Decision Theory, Probability

## 1. Introduction

The set-up of the two envelope paradox has been discussed extensively in the literature (Nalebuff 1989; Broome 1995; Chalmers 2002), so I shall be brief in that regard. You are shown two envelopes, both containing an unknown amount of money, and told that one has twice as much as the other. Upon selecting one of the envelopes you are given the choice to switch envelopes. Should you? Set $X$ equal to the amount in the envelope you currently have. There is a .5 probability the other envelope contains $.5X$ and a .5 probability that it contains $2X$. It seems to follow from Bayesian decision theory that the other envelope is worth $.5(.5X) + .5(2X)$, or $1.25X$, making it favorable to switch. This is obviously impossible as we could have used the same reasoning had we chosen the other envelope.

In 1996, Bruss presented what at first appears to be a solution to the two envelope paradox, and indeed it dispels this version of the paradox successfully. He explains that the paradox arises because $X$ cannot be treated as a constant. If after being offered the choice of envelopes we reason that the other envelope has a .5 probability of containing .5X, we have assumed X is the value of the larger envelope. However, when we reason that there is a .5 probability that the other envelope will contain 2X, X must represent the value of the smaller envelope. Consequently we cannot add those terms together as X does not represent the same quantity in both terms (Bruss 96).

However, the paradox is not entirely eliminated. Suppose we had opened the envelope and then been offered the opportunity to switch. Once we observe that the envelope contains, say, $10, that amount obviously is a constant. One cannot very well argue that $10 has two separate values. Thus, the same mathematical logic used by Bruss would no longer be applicable, and the paradox is reestablished. So far no author has suggested a response to this complication that has gained significant traction.

One popular response advocated by Christensen and Utts (1992) and Nickerson and Falk (2006) among others, is to reject the idea that there is a subtle flaw in the mathematics and instead suggest one choose the envelopes using prior knowledge of how much money is available. The idea is that if one can make assumptions about how much money is being used, one can calculate the probability distribution from which the amounts are chosen and thus make an educated decision on whether or not to switch envelopes.

While this methodology would indeed be practical if one did have prior knowledge, this is not a legitimate solution. First, it is trivial to imagine a scenario in which one has no knowledge of the distribution. Second, and even more importantly, it evades the real issue. It is akin to attempting to explain the Monty Hall problem by arguing that one should simply look at the host's facial expressions for clues as to which door to pick.

## 2. A New Perspective

To see how the paradox can be resolved in its entirety, we must examine it from a different perspective. Previously the literature has focused solely on the perspective of the player who is offered the envelope, not that of the offering party. To rectify that, let us set the game up where we are offering the money. Let us suppose we know that we have two envelopes, one containing $5.00 and one containing $10.00. We offer the player the knowledge that one envelope has twice as much money as the other and the opportunity to pick one.

If the player picks the $5.00 envelope, is allowed to look inside, and then is given the opportunity to switch he could reason he has a .5 probability of losing $2.50 and a .5 probability of gaining $5.00; obviously, he reasons, it is in his best interest to switch. If there was a third party who knew the values of the envelopes but did not observe which amount the player chose, he would conclude that there was a .5 chance that switching would gain him $5.00 and a .5 chance that it would lose him $5.00. This third party would also know that both envelopes have a value under Bayesian probability of $7.50; clearly, there is no value in switching. We know that there is in fact a 0.0 probability of his losing $2.50 and a 1.0 probability of his gaining $5.00; therefore, we know that he should of course switch.

The player however, has estimated the relative utility of his options as well as he might be expected based on the probabilities he calculates from his subjective position and the use of Bayesian decision theory. In this case, even though his information was incomplete, his estimation based on subjective probability seemed to be helpful; he chose the larger amount of money, but because his reasoning was flawed it is only by luck on his part that he chose the more favorable option.

If, on the other hand, the player picks the $10.00 envelope, under the same conditions, he could reason he has a .5 probability of losing $5.00 and a .5 probability of gaining $10.00. Should he not then take the other envelope? By his estimate, it is clear he should. If again there were a third party who knew the values of the envelopes but not which one the player chose, he would arrive at the same conclusion as before, that there was a .5 chance that switching would gain him $5.00 and a .5 chance that it would lose him $5.00. We, on the other hand, know that in fact his odds off losing $5.00 are 1.0 and his odds of gaining $10.00 are 0.0.

In this case, while he still calculates the Bayesian interpretation as well as can be expected with the information afforded to him, his conclusion that he should switch was less advantageous than it would have been had he chosen the other envelope first. In essence, Bayesian decision theory was not helpful.

### 3. The Source of the Paradox

The reason the Two Envelope problem appears so problematic is that Bayesian Decision theory does not, on average, give good advice to the player using their subjective estimations of the probabilities of different outcomes. The third party who knew the values of the envelopes correctly applied Bayesian decision theory to determine that there was no rational reason to switch without further information. We, the controllers of the game, who had that further information, obviously knew whether or not the player should switch.

The player, however, had to rely on assumptions that, while reasonable, were wrong. If the player chose the envelope with $10.00, he believed there to be a .5 chance that the other envelope contained a $20.00 payoff. While that assumption cannot be faulted based on the information he had, it simply was wrong. There was no possibility of a $20.00 payoff. It skewed his value expectations higher. Had $20.00 been an option, had the value of the second envelope been determined by a coin toss between $5.00 and $20.00 after opening the first, it would have been in the player's best interest to switch. The erroneous belief in the possibility of a $20.00 payoff creates the problem with subjective probability.

The hypothetical player who was fortunate enough to pick the $5.00 envelope first used the same problematic reasoning. It was simply his good luck that in fact the other envelope did contain $10.00. He was mistaken about the possibility of losing $2.50, but the essential element is that a player who expects the possibility of a lower payoff that doesn't exist is off by less than a player who expects the possibility of a higher payoff. This disparity is the reason the Bayesian interpretation is skewed too high, leading the player to believe he should switch.

Just as Bruss demonstrated why we cannot assign a single variable to represent the value of the first envelope, this reasoning shows what goes wrong when the game itself tells us the value of first envelope, forcing us to have a single definite value.

**4. The Mathematics behind the Paradox**

If this theory is correct, we would expect that it would be able to predict the factor of 1.25 by which one naively might believe switching would be better than not switching, and in fact it is able to do so. Consider two envelopes with $X$ and $2X$ dollars. There is a .5 chance that one will pick the envelope with $X$ dollars, in which case he will believe there is a .5 chance that the other envelope has $.5X$ dollars and a .5 chance that it has $2X$ dollars. He will thus conclude that the expected value of the other envelope is .5(.5X+2X), or 1.25X dollars. There is a .5 chance that one will pick the envelope with 2X dollars, in which case he will believe there is a .5 chance that the other envelope will have X dollars and a .5 chance that it has 4X dollars. He will thus conclude that the expected value of the other envelope is .5(X+4X), or 2.5X dollars.

Combining those values we obtain .5(1.25X + 2.5X) or 1.875X. This represents the value that, on average, a player would attribute to the other envelope in terms of the ratio of it to the smaller envelope. In comparison, a third party knowing the values of the two envelopes would regard the value of each envelope on average to be .5(x+2x) or 1.5X. This represents the value that, on average, such a third party would attribute to the other envelope in terms of the ratio of it to the smaller envelope. The ratio by which the first estimate is off from the estimation of the objective third party is, as expected, 1.875X/1.5X or 1.25 (eliminating the term X). The resultant quantity, 1.25, represents how much greater than the objective value of the other envelope a player would, on average, expect it to be. Clearly, now, we can see how the faulty assessment of probability leads precisely to the erroneous belief that the other envelope's Bayesian value is 1.25 times greater than the first.

## 5. Conclusion

Obviously the player is not in a position in which it is impossible to recognize that there is no rational reason to switch. On the contrary, a perfectly rational player would simply recognize that his subjective probabilities provide a misleading account using Bayesian decision theory and would therefore ignore those results. As Falk and Nickerson noted in 2009, "Without seeing the value in one's envelope, one should be indifferent." Now it is clear why that indifference should stand when one observes the contents of an envelope.